\documentclass[conference]{IEEEtran}

\ifCLASSINFOpdf
  \usepackage[pdftex]{graphicx}
\else
\fi

\usepackage[T1]{fontenc}
\usepackage[english]{babel}
\usepackage{algorithm,algorithmicx,algpseudocode}
\usepackage{amssymb}
\usepackage{multicol}

\begin{document}

\title{MAMoC: Multisite Adaptive Offloading Framework for Mobile Cloud Applications}

\author{\IEEEauthorblockN{Dawand Sulaiman}
\IEEEauthorblockA{School of Computer Science\\
University of St Andrews\\
St Andrews, UK\\
djs21@st-andrews.ac.uk}
\and
\IEEEauthorblockN{Adam Barker}
\IEEEauthorblockA{School of Computer Science\\
University of St Andrews\\
St Andrews, UK\\
adam.barker@st-andrews.ac.uk}}

\maketitle

\begin{abstract}
This paper presents MAMoC, a framework which brings together a diverse range of infrastructure types including mobile devices, cloudlets, and remote cloud resources under one unified API. MAMoC allows mobile applications to leverage the power of multiple offloading destinations. MAMoC's intelligent offloading decision engine adapts to the contextual changes in this heterogeneous environment, in order to reduce the overall runtime for both single-site and multi-site offloading scenarios. MAMoC is evaluated through a set of offloading experiments, which evaluate the performance of our offloading decision engine. The results show that offloading computation using our framework can reduce the overall task completion time for both single-site and multi-site offloading scenarios.
\end{abstract}

\IEEEpeerreviewmaketitle

\section{Introduction}

With the sheer ubiquity of powerful mobile devices, it is now possible to harness the capabilities of these mobile devices in order to execute, and offload compute-intensive applications. Most of the modern mobile devices on the market have quad-core and octa-core CPUs, and mobile devices of today are more computationally powerful than the PCs of the last decade \cite{6726211}. 

%Nonetheless, mobile devices are still limited in computation power and battery lifetime with respect to their fixed counterparts (e.g., modern desktop computers, laptops, and data center servers).

The concept of using cloud hosted infrastructure as a means to overcome the resource-constraints of mobile devices is known as \emph{Mobile Cloud Computing (MCC)}, and allows applications to run partially on the device, and partially on a remote cloud instance, thereby overcoming any device-specific resource constraints. However, as smart phones and tablets gain more CPU power and longer battery life, the meaning of MCC gradually changes. Instead of being fully dependent on the cloud, a number of nearby devices can be used to coordinate and distribute content and resources in a decentralized manner; this is known as \emph{Mobile Ad hoc Cloud Computing} \cite{WCM:WCM2709}.  

%Even though mobile computing has advanced its state but the energy and computational resources are still limited.

Although MMC has helped many application developers to overcome the limited resources of mobile devices, it has also created a new set of challenges, such as the possibility of high network latencies and low bandwidth availability between the mobile device and the cloud. Other research efforts investigated frameworks which allow the cloud to move closer to the user in the form of cloudlets \cite{5280678} \cite{Verbelen:2012:CBC:2307849.2307858}. Cloudlets are trusted, resource-rich computers that are connected to the Internet and available to nearby mobile devices. Other approaches include a group of nearby mobile devices to leverage lower end devices thus the formation of a local mobile cloud \cite{Shi:2012:SER:2248371.2248394} also referred as \emph{Mobile Device Clouds} \cite{6753815} and \emph{Mobile Edge Clouds} \cite{fernando2016computing}. Mobile devices with less computational power and lower battery life can be empowered by the nearby mobile devices to run resource-intensive applications. Therefore, more efficient and reliable methodologies need to be explored for resource-hungry and real-time applications such as face recognition, data-intensive, and augmented reality mobile applications. 

The heterogeneous mobile cloud environment contains different types of computing resources such as remote clouds, cloudlets, and mobile devices in the vicinity that can be utilized to offload mobile tasks. Heterogeneity in mobile devices includes software, hardware, and technology variations. In MCC, providing collaboration among various mobile and cloud nodes with different interfaces is a significant matter. Dynamic environmental changes is one of the challenges facing the offloading decision making in mobile cloud applications. Mobile Cloud frameworks need to adapt to these changes for efficient task partitioning and high QoS of the mobile applications running on end user devices. Existing mobile computation offloading frameworks lack the automated transparency feature so that the surrounding devices can be detected and the computation offloading take place in a seamless manner \cite{sanaei2014heterogeneity}. 

This paper presents \textit{Multisite Adaptive Mobile Cloud (MAMoC)}, a unified framework which allows each mobile device within the shared environment to intelligently offload its computation to other external platforms. For the individual mobile devices, it is important to make the offloading decision based on network conditions, load of other machines, and mobile device's own constraints (e.g., mobility and battery). Moreover, to achieve a global optimal task completion time for tasks from all the mobile devices, it is necessary to devise a task scheduling solution that schedules offloaded tasks in real time. To achieve the vision of mobile computing among heterogeneous converged networks and computing devices, designing resource-efficient environment-aware applications is an essential need. The offloading decision engine needs to adapt to the dynamic changes in both the host device and connected nearby and remote devices. Changes in hardware such as CPU workload, available memory, and battery state and level need to be carefully monitored. 

The main contributions of this work are:

\begin{enumerate}

\item Leveraging mobile applications running on lower end devices with nearby mobile devices, cloudlets, and remote cloud resources using a single framework.

\item An API on top of our framework, which allows application developers to use it as a simple programming model to build mobile cloud applications and abstract complex underlying heterogeneous technologies.

%\item Using edge devices (routers, switches, servers) in the computation offloading process. Using lightweight containers help achieving this goal.

\item Through a real-world evaluation, we demonstrate that it is possible to speedup computation of mobile devices by using the adaptive offloading algorithm in both single-site and multi-site offloading scenarios.

\end{enumerate}

%The development of a unified and comprehensive mobile cloud system remains a challenging task. Mobile Cloud Computing implies that mobile application developers must learn how to make use of cloud computing, which can in turn entail financial burdens. An area of interest in MCC literature is to support application developers in creating such component based applications and defining their quality constraints. Splitting up mobile applications in a (semi-) automatic way decreases the development effort and time \cite{Verbelen:2012:CBC:2307849.2307858} \textbf{This needs to be rewritten}

%Making offloading decisions can be described as a task partitioning problem which is NP-complete. Compared with the traditional cloud computing, MCC has its own challenges including mobility, diversity of network conditions and limited channel bandwidth. The changes occurring during applications' life cycle need an adaptive runtime task repartitioning. Most existing works focus on the static single site MCC with an assumption that wireless network conditions are stable and there is only one server. Some works which study the dynamic MCC aim to optimize the consumption of individual components rather than the whole application. In this paper, we focus on multisite computation offloading in dynamic mobile cloud environments with the consideration of environmental changes during the task execution. Multisite computation offloading can migrate components to multiple service providers. \textbf{This takes too long to get to the point}

\section{Related Work}

Mobile Computation offloading transfers processing from the mobile device to other service providers. Mobile application is partitioned and analyzed so that the most computational expensive operations at code level can be identified and offloaded \cite{Kumar:2013:SCO:2431282.2431342}. The objective is to improve the computation performance, enable advanced functionality, and preserve scarce resources. In the mobile-to-cloud offloading model, most challenges arise from partitioning the mobile application code to remote and local tasks based on the dependencies of each task. As such, the current solutions can be categorized by partitioning technique into static and dynamic. The authors at \cite{Zhang2011} produce an elastic application model in the form of weblets which can be platform independent or dependent. The decision of offloading is based on contextual components stored in the cloud including device status (CPU load, battery level), performance measures of the application for quality of experience and user preferences. By doing so, the application model supports multiple running modes: power-saving mode, high speed mode, low cost mode or offline mode. Mobile offloading overcomes the resource limitations of lower end devices by splitting resource-intensive tasks and allocating subtasks to other resource-rich devices. Offloading may be performed at different granularities ranging from methods and individual tasks \cite{Zhang2011} to applications \cite{Giurgiu:2009} and virtual machines \cite{Chun:2011:CEE:1966445.1966473}.

The mobile cloud framework developed in \cite{Huerta-Canepa:2010:VCC:1810931.1810937} uses the same interface of existing cloud APIs for the collocated virtual computing providers. This allows a seamless integration with the existing cloud infrastructures. On the other hand, Cuckoo framework \cite{kemp2010cuckoo} uses the native Android partitioning to separate the user interface from the background computational code. This eases the design and implementation of MCC applications, as mobile application developers do not require any cloud computing knowledge, such as integrating with offloading APIs. 
%Adaptive offloading mechanisms and context awareness offloading schemes are studied in the Mobile Cloud Computing literature. \cite{6533324}  mCloud Bowen zhou XXXXX

%Authors at \cite{6726211} study the case of mobile edge clouds also called ad hoc mobile clouds. Even though the usage of the term has slightly changed after the introduction of Edge Computing. In the literature, Mobile Edge Computing is either used as near network device data processing \cite{} \cite{} or offloading to nearby mobile devices. Our system scenario takes mobile edge clouds as a collection of nearby mobile and immobile devices with a part of the devices connected to the remote cloud if further processing power and data is needed.

Most existing MCC proposals concentrate on single-site offloading \cite{Cuervo:2010:MMS:1814433.1814441} i.e., offloading application's parts from the mobile device to a single server. However, as the number of surrounding devices and cloud computing and storage increases, it is more common that an application can be executed on multiple servers \cite{Sinha2011TechniquesFF}. It is shown that we can obtain better performance from multisite offloading. Therefore, multisite offloading is considered as a generally realistic model in this work. However, making decision for multisite offloading problem is an NP-complete problem, and hence, obtaining the optimal solution is time consuming. Hence, we use a simple near optimal decision algorithm to find the best-possible partitioning for offloading to multisite clouds/servers.

We have shown in our previous work that offloading tasks from lower end devices to nearby devices can save up to 50\% in computation performance \cite{sulaiman2016task}. In this work, we propose the creation of a hybrid mobile cloud framework in which the resources are nearby mobile, immobile (cloudlets) and remote cloud servers. MAMoC focuses on reducing the total task computation time and the overall consumed power of the interconnected devices. Similar to \cite{kemp2010cuckoo}, we rule in favor of injecting the offloading code transparently to the developer. By doing so, we preserve the mobile application development process, as the developer is unaware of the underlying mechanisms and is only left with the possibility of giving hints about code that may be offloaded. Such a development model encourages the decoupling of components and the modularity of mobile applications.
%In this work, we expand the idea to allow the offloading process to happen on other fixed nearby devices (cloudlets) as well as remote clouds in case of the presence of high bandwidth connectivity.

\begin{figure}[!t]
\centering
  \includegraphics[width=\linewidth]{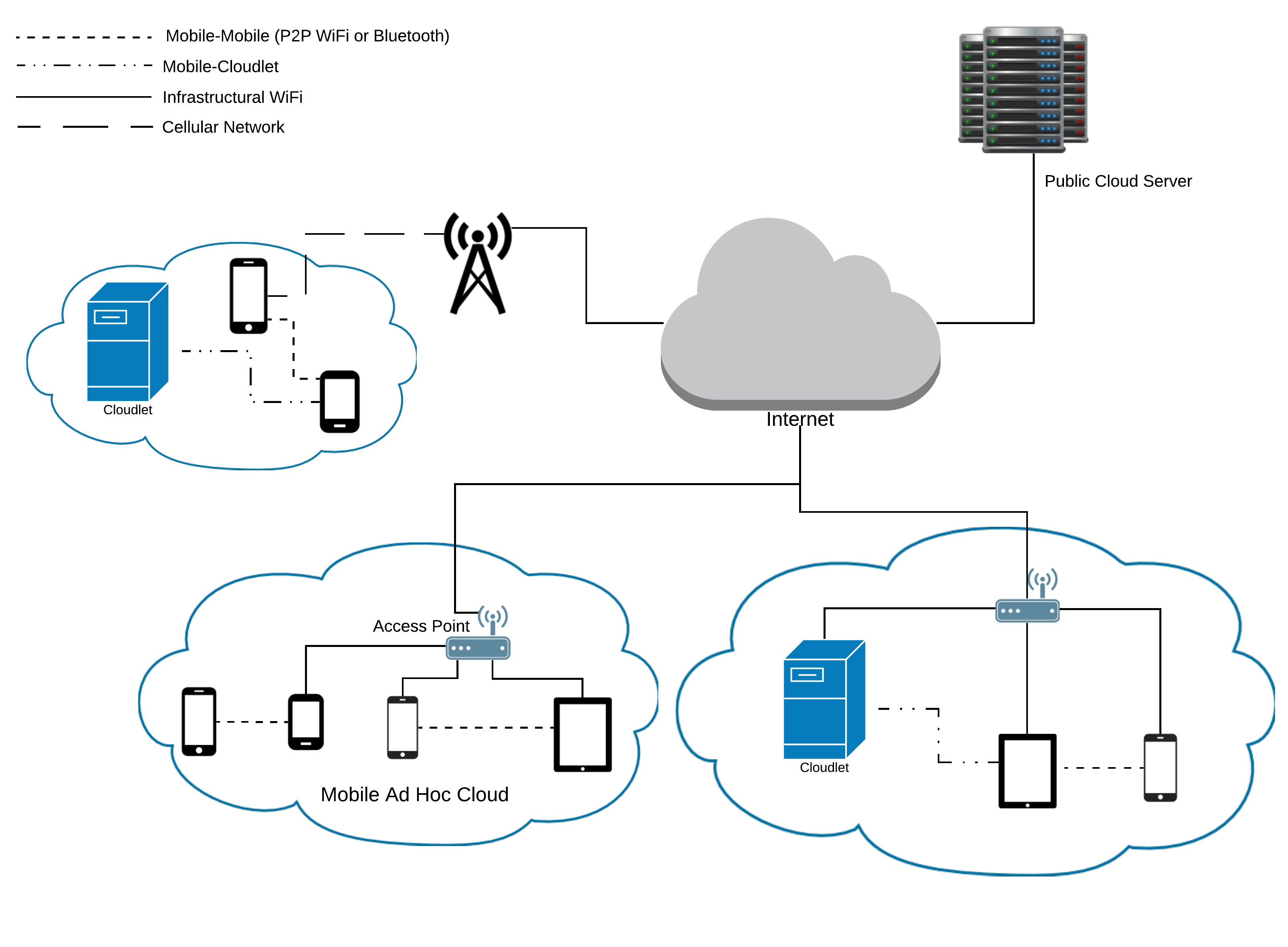}
  \caption{Heterogeneous Multisite MCC Architecture}
  \label{MCC-Architecture}
\end{figure}

\section{System Design}

Our objectives for MAMoC include improving the running time of the compute-intensive tasks on the mobile devices as well as saving energy and bandwidth by computation offloading. MAMoC is designed to allow mobile devices to discover other surrounding devices over standard Local Area Network in infrastructural Wi-Fi using an access point, peer-to-peer Wi-Fi, and Bluetooth personal area networks as shown in Figure \ref{MCC-Architecture}. The main goals of the framework are to allow mobile application developers to achieve a transparent automated offloading to multiple destination clouds (mobile clouds and public clouds) and device dynamic changes over the lifecycle of execution of an application. 

\subsection{Components}

\subsubsection{Service Discovery} 
\begin{figure}[!t]
\centering
  \includegraphics[width=2.5in]{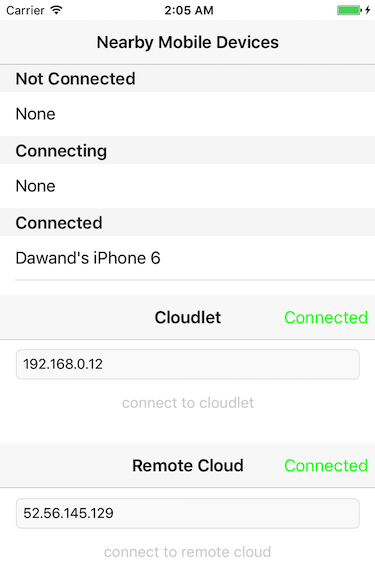}
  \caption{Service Discovery}
  \label{service_discovery}
\end{figure}

After the framework is initialized, service discovery is performed. Each mobile device can advertise services and discover what services other nearby devices on the local network are offering. A browser object in a host device searches for peers which have an advertiser object. This can be done using infrastructural Wi-Fi, where the devices are connected to the same Access Point, Peer-to-peer Wi-Fi, or Bluetooth wireless technologies. An offloader device can also scan the local network for cloudlet servers and open a TCP socket communication channel for offloading requests. A standard user interface is developed to be used by the applications for service discovery and managing device connections as shown in Figure \ref{service_discovery}.

The mobile ad hoc cloud communication technique over LAN is based on Zero Configuration Network \cite{cheshire2006zero}. MultiPeer Connectivity (MPC) library \cite{applempc} uses Zero Configuration Network technology to allow application developers to form short range sessions between nearby devices. The devices do not need a preconfigured network to exchange data with each other. There is no configuration needed because they can discover each other via multicast DNS (mDNS) \cite{cheshire2013multicast}. mDNS is a service that resolves host names on a local network without the use of a central domain name server. After the connection is established between two devices, data can be exchanged between them in either reliable mode over TCP or unreliable mode over UDP. The reliable mode guarantees the delivery and avoids out of order packages so we have used this mode throughout our framework data transfers.
%Instead a resolving host simply sends a DNS query to a local multicast address and the host with that name responds with a multicast message with its IP address. Multicast DNS is also used in combination with DNS based service discovery where a host that provides a network service can announce its service to the local network. Those services can then be discovered using multicast messages. 

\subsubsection{Device Profiler} This component collects real time information of the devices including hardware, software, and networking related context information. This allows the framework to have sufficient information about the connected devices. This eases the process of offloading decision making. Once a new device joins the framework, it goes through two profiling procedures. First, a collection of contextual information of the device is retrieved by the framework including hardware information such as number of CPU cores and processing speed, the type of the network (Wi-Fi or cellular), and battery level and state (whether it is charging or not). Second, a workload is sent over to the device after a successful connection establishment. The makespan result of running the workload is sent back to the framework. The collected information in both phases is then disseminated to all the connected devices in the framework to be used by the decision engine in the offloading process.

\subsubsection{Offloading Decision Engine} The engine is fed with the metrics from device profiler component and it is used by the framework to check whether to offload a task to other devices or execute it locally. This component is only used in the partial offloading execution modes. Unlike full offloading, the framework needs to decide the candidate destinations for offloading the task, partition the task into a number of subtasks, and distribute them.

\subsection{Offloading Models and Algorithms}

When we model mobile computation offloading problems, we need to consider the time needed to execute computation locally, time needed to execute computation remotely (excluding the cost of data transfer), and time needed to move input data and results back and forth between the local and remote computers. We also need to collect the device context information in our preprocessing phase to model the offloading decision making process including the CPU power, memory, battery state and level of the connected devices.

\begin{table}[!t]
\centering
\caption{Notations}
\label{notations}
\begin{tabular}{ll}
\cline{1-2}
{\textbf{Symbol}} & {\textbf{Description}}  \\ \cline{1-2}
$ B_{man} $ & Benchmark workload result of mandelbrot set  \\
$ B_{ftt} $ & Benchmark workload result of FFT  \\
$ B $ & Average Benchmark workload result \\
$ B_{mob} $ & Benchmark score of a mobile device  \\
$ B_{clet} $ & Benchmark score of a cloudlet  \\
$ B_{c} $ & Benchmark score of a remote cloud  \\
$ BL_{mob} $ & Battery level of a mobile device \\
$ M_{mob} $ & Available memory of a mobile device  \\
$ M_{clet} $ & Available memory of a cloudlet  \\
$ M_{c} $ & Available memory of a remote cloud  \\
$ OS_{mob} $ & Offloading score of a mobile device  \\
$ OS_{clet} $ & Offloading score of a cloudlet  \\
$ OS_{c} $ & Offloading score of a remote cloud  \\
$ P_{mob} $ & Computation speed of a mobile device  \\
$ P_{clet} $ & Computation speed  of a cloudlet  \\
$ P_{c} $ & Computation speed  of a remote cloud  \\
$ RTT_{mob} $ & Network overhead and data transfer to mobile device  \\
$ RTT_{clet} $ & Network overhead and data transfer to cloudlet  \\
$ RTT_{c} $ & Network overhead and data transfer to cloud  \\
$ TOS $ & Total offloading score of all the connected nodes \\
\cline{1-2}
\end{tabular}
\end{table}

The benchmark score will be calculated after the workloads are sent to the devices and the execution results are received. The quicker the CPU completes the tests, the higher the benchmark score. The workloads measure the instruction performance of the device by performing processor-intensive tasks that make heavy use of integer instructions. Initially, we create two types of workloads: compute-bound and memory-bound. Mandelbrot set \cite{fernando2016computing} of an 800x800 pixels is used for the first type. The Fast Fourier Transform (FFT) \cite{katoh2002mafft} is used as a memory heavy workload. The workloads are executed 5 times and the average runtime score in GFlops is returned to MAMoC. Finally, the Benchmarking score of the device is calculated using equation \ref{equation1}:

\begin{equation}
B = (B_{man} + B_{ftt}) / 2
\label{equation1}
\end{equation}

The computation power and memory of connected nearby mobile devices are initially collected. If the mobile device is not currently charging, then the offloading score is deducted by the amount of the used battery level of the device. The offloading score is then calculated using the following equation:

\begin{equation}
OS_{mob} = (B_{mob} + P_{mob} + M_{mob}) - RTT_{mob}  - (100 - BL_{mob}) 
\end{equation}

Similar to mobile ad hoc cloud model, cloudlet and remote cloud modelling is a summation of their respective benchmark score, computation speed, and available memory subtracted by the data transfer cost.

\begin{equation} OS_{clet} = (B_{clet} + P_{clet} + M_{clet}) - RTT_{clet} \end{equation}

% We assume the communication overhead of nearby devices (both mobile devices and cloudlets) to be negligible. However, we need to consider it for our cloud modelling. 

\begin{equation} OS_{c} = (B_{c} + P_{c} + M_{c}) - RTT_{c} \end{equation}

Given the offloading scores for all the offloadee device candidates, we calculate the total offloading score for any particular task execution.

\begin{equation}
TOS = |OS_{mob}| + |OS_{clet}| + |OS_{c}|
\end{equation}

MAMoC collects the offloading scores of the local device running the mobile application and all the connected service providers. Algorithm 1 shows the process of collecting individual offloading scores calculated and received earlier to generate a dictionary of nodes and their corresponding offloading scores.

\begin{algorithm}[H]
 \caption{Offloading Score Algorithm}
 \begin{algorithmic}[1]
 \renewcommand{\algorithmicrequire}{\textbf{Input:}}
 \renewcommand{\algorithmicensure}{\textbf{Output:}}
 \Require connectedNodes
 \Ensure  nodeScores
  \State nodeScores = [:] \Comment{A dictionary of nodes and their respective offloading scores}
 \State selfNode = getSelfNode()
 \State localScore = getScore() \Comment{Based on equation 2}
 \State nodeScores.add(selfNode, localScore)
 \If {connectedNodes \textgreater 1}
 \For{node in connectedNodes}
            \State score = node.getScore() \Comment{Based on equations 2-4}
            \State nodeScores.add(node,score)
      \EndFor
   \EndIf \\
   \Return nodeScores
 \end{algorithmic} 
\end{algorithm}

The node scores will then be sent to task partitioning algorithm to calculate the final task partitioning percentage for any given task. 

\begin{algorithm}[H]
 \caption{Task Partitioning Algorithm}
 \begin{algorithmic}[1]
 \renewcommand{\algorithmicrequire}{\textbf{Input:}}
 \renewcommand{\algorithmicensure}{\textbf{Output:}}
 \Require nodeScores, totalScore
 \Ensure  partitioningResult
 \State partitioningResult = [:] \Comment{A dictionary of nodes and task allocation percentages}
 \For{(node,score) in nodeScores}
     \State partition = (score / totalScore) * 100
     \State partitioningResult.add(node,partition)
 \EndFor \\  
   \Return partitioningResult
 \end{algorithmic} 
\end{algorithm}

\subsection{Development Phase}

We have used Swift \footnote{https://swift.org} to implement both the client (iOS) and server (Linux) components of MAMoC. Swift is open sourced by Apple on December 2016. This has enabled developers to write Swift applications on broader range of platforms. To the best of our knowledge, this is the first complete Swift and iOS based framework in the mobile cloud computing research literature.

Containers are utilized to host the server components of MAMoC. Linux Containers (LXC) is a virtualization method for running multiple isolated Linux systems on a single machine. Docker\footnote{https://www.docker.com/} extends LXC to automate the deployment of applications inside software containers. We have used an already developed Swift Docker image to implement a server-side swift application to accept incoming requests from the mobile devices. The developed container provides an environment which is ready to be customised for other mobile applications. It provides a feature-rich yet lightweight execution environment for offloaded tasks.

\begin{figure}[!t]
\centering
  \includegraphics[width=3.5in]{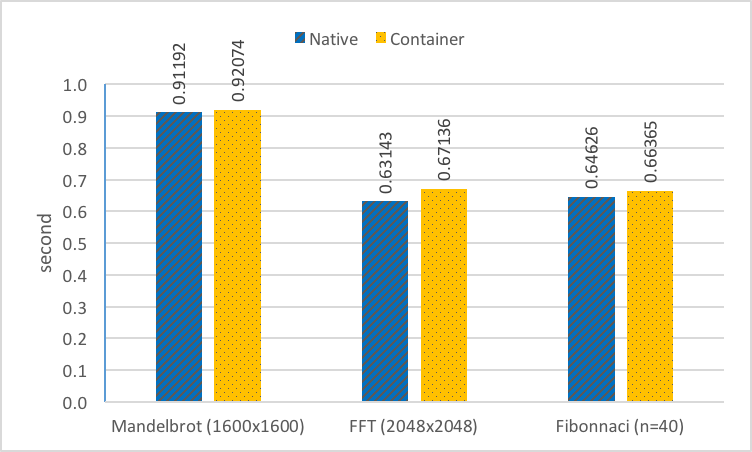}
  \caption{Container vs. native benchmarking}
  \label{containerBenchmarking}
\end{figure}

We have run a couple of compute-intensive workloads to experiment the speed of computation on Docker containers as well as on native platforms. As the results show in \ref{containerBenchmarking}, the computation speed of containers are close to native platform performance within 1\% region \cite{7095802} \cite{7484184}. To ease the creation of the container, we provide a Dockerfile that can automate the process of creating a container for the server-side part of MAMoC. The container is preconfigured with the necessary build environment for handling the client requests.

MAMoC allows other mobile application developers to adapt it for their own compute-intensive mobile applications. A developer must follow the following steps to configure an application to use our framework:
\begin{enumerate}
\item Add MAMoC to the mobile application: This process can be done manually by adding the files to the application directory or automatically using CocoaPods\footnote{https://cocoapods.org/} which is a dependency manager for Swift and Objective-C projects.

\item Define the task and its parameters: In this step, the developer has to override \textit{initTask} method to provide any user input or additional data to be processed by the framework. In our text search example, the parameters are the text file and the search keyword which is entered by the user. In a face recognition application, the parameter would be an image of the face and so on.
 
\item Define the offloading sites: Even though MAMoC is designed to offload tasks to multiple sites by default, the developer can define where to execute the task. There are four choices to choose from: nearby mobile devices, cloudlet servers (local immobile), remote cloud servers, or automatic. The automatic option depends on the outcome of the models in the device profiler component and how the tasks are partitioned to target offloadees by the offloading decision engine.
 
\item Execute the job: the subtasks will be distributed accordingly to be processed in parallel on the target offloadees. The results are returned and merged together in the offloader device and presented to the user.

\item Set the timeout for the task reprocessing (optional): For the tasks which are not returned successfully by the resource providers, the task will be reprocessed locally in the device once the timeout duration is reached. The default value for this is 10 seconds. This adds fault tolerance feature to the applications developed on top of MAMoC.

\end{enumerate}

The source code of MAMoC and a short documentation for setting up the different components in the framework is publicly available online at \textsl{https://github.com/dawand/MAMoC}

\section{Experimentation}

\begin{table}[!t]
\centering
\caption{Testbed Experimentation}
\label{testbed}
\begin{tabular}{|c|c|c|c|}
\cline{1-4}
{\textbf{Node}} & {\textbf{Benchmark}} & {\textbf{CPU}} & {\textbf{RAM}} \\ 
 & {\textbf{(in GFlops)}} & {\textbf{(in GHz)}} & {\textbf{(in GB)}} \\ \cline{1-4}
 Mobile (small) & 1.09 & 1.3 (Dual) &  1  \\ \cline{1-4}
 Mobile (medium) & 1.24 & 1.4 (Dual) &  1  \\ \cline{1-4}
 Cloudlet & 2.56 &  2.5 (Quad) &  16  \\ \cline{1-4}
 Cloud (small) & 2.32 &  2.4 (Single) &  1  \\ \cline{1-4}
 Cloud (medium) & 2.94 &  2.8 (Quad) &  7.5  \\ \cline{1-4}
 Cloud (large) & 3.02 &  2.8 (Octa) &  15  \\ \cline{1-4}
\end{tabular}
\end{table}

%The test plan revolves around a prototype application that is deployed on the mobile devices. 
We have developed a mobile application to test the performance of the framework and showcase the different execution scenarios. We measure total completion time for the application in different settings. The application is executed in four different modes: local execution, nearby mobile devices, cloudlet, and remote cloud (with three different servers). We use four offloading scenarios: a full offloading and three types of partial offloading (workload sharing in a parallel manner) with different configurations. The application contains a Knuth-Morris-Pratt searching algorithm \cite{knuth1977fast} to be performed on three different size text files. The large text file consists of 1,095,649 words, the medium text file contains 316,323 words, and the small text file contains 39,799 words. The files are stored in the mobile application running on the offloader device. The files need to be distributed through a wireless medium to the offloadees before the keyword search is conducted in the destination. Each execution was repeated ten times, such that averages could be calculated for more accurate results.

Our experimental testbed consists of two mobile devices (an iPhone 5 and an iPhone 6), one cloudlet, and three remote cloud instance types. The hardware specifications are shown in Table \ref{testbed}. For the remote cloud instances, we have used three different Amazon Web Services instance types: small (t2.micro), medium (c4.xLarge), and large (c3.2xlarge). The two phones will be running the mobile application while cloudlet and remote cloud instances will be running the container described earlier.

In all our experiments, mobile (small) is the offloader and the rest of the nodes are service providers (offloadees). The offloading scenarios are:

\begin{itemize}
\item Full offloading: In full offloading mode, the execution is performed in the offloadee and the final result is returned to the offloader. On the other hand, partial offloading mode only sends a part of the execution over to the offloadee and performs the rest of the execution itself as we will observe in the other two scenarios. After the results are received, they are merged and stored in the offloader. 

\begin{figure}
\centering
\label{fulloffloading}
  \includegraphics[width=\linewidth,height=2in]{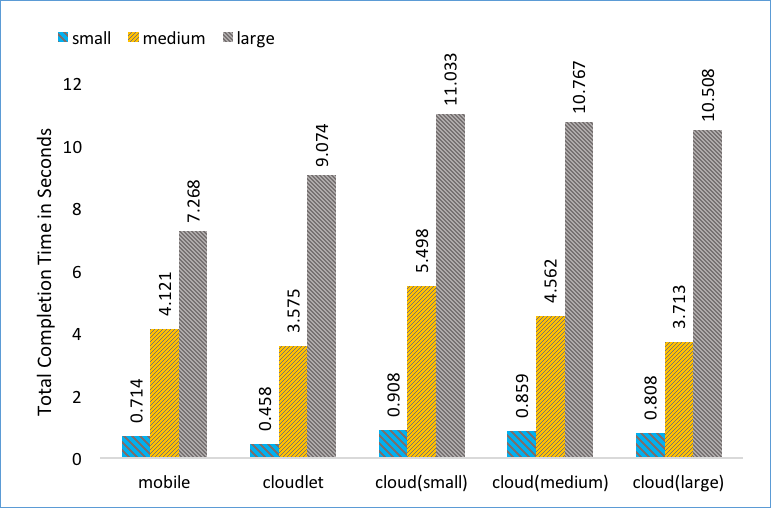}
  \caption{Full Offloading - The whole computation is offloaded to the offloadee}
\end{figure}

Our framework is specifically designed to support partial offloading. Nonetheless, we wanted to observe the completion times of full offloading executions and compare them to the decision engine results. We performed a full offloading experimentation on the offloadees separately using the three text files. Figure 4 shows the total completion time of running the application with different text file sizes. As we will see later, most of the completion time is the communication overhead that occurs during the transfer of the necessary data (the text file content) from the mobile device to the offloadees.

\item Partial offloading: we perform our first set of partial offloading experiments with no help from MAMoC. The tasks are equally distributed among the connected nodes. In other words, if there is only one available offloadee, the workload is divided into two equivalent halves and distributed to them. Both devices then execute the workload in a parallel fashion. The local result and the result returned from the offloadee are then merged and stored locally. The results of running the same set of workloads as previous experiment are displayed in Figure 5.
%\ref{partialoffloading5050}.

\begin{figure}
\centering
\label{partialoffloading5050}
  \includegraphics[width=\linewidth,height=2in]{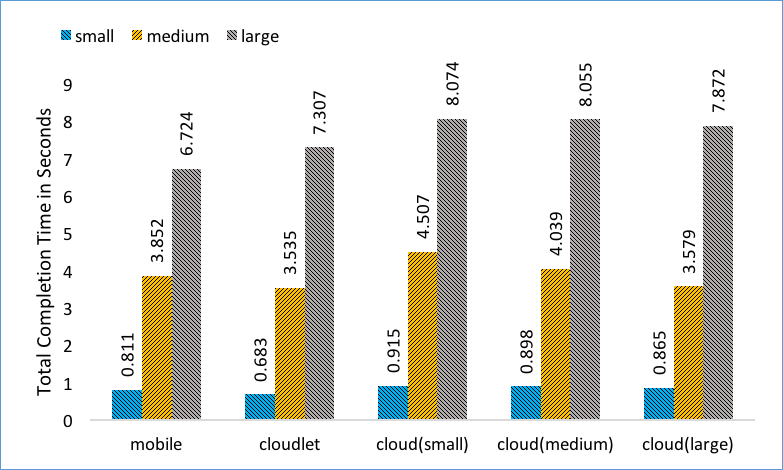}
  \caption{Partial Offloading (equal task distribution) - Local mobile device executes 50\% of the task while the remaining 50\% is offloaded}
\end{figure}

\item Partial offloading with decision engine enabled: the decision engine uses the models presented earlier to calculate the percentage of the task which should be offloaded to any particular offloadee. The offloading scores of the offloader and the offloadees are investigated by the decision engine for the task partitioning process. In the case of offloading score being less than zero, no computation is offloaded to that offloadee. We wanted to observe the performance of our decision engine in both single-site and multi-site offloading scenarios. Figure 6 shows the single-site offloading scenario where the task is meant to be executed locally and a single offloadee. The task partitioning percentage and completion times of partitioned tasks for multi-site offloading scenario are displayed in Figure 7. It is worth noting that we have only used the large instance type of the remote cloud along with a cloudlet and a nearby mobile device as offloadees in multi-site offloading scenario. The complete set of results of all the experiments are shown in Table III.

\end{itemize}

\begin{figure}
\centering
\label{partialoffloadingsingle}
  \includegraphics[width=\linewidth,height=2in]{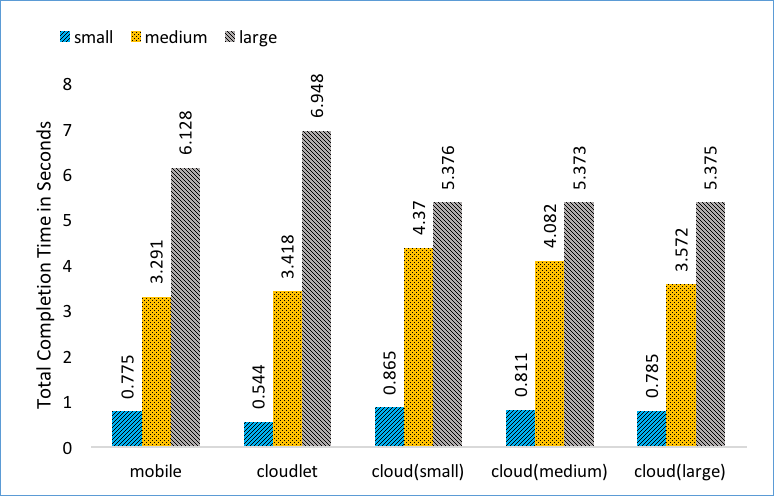}
  \caption{Partial Offloading with MAMoC (single-site) - The decision engine partitions the task for the local mobile device and a single offloadee}
\end{figure}

\begin{figure*}
\label{partialoffloadingmulti}
\begin{multicols}{2}
    \includegraphics[width=\linewidth]{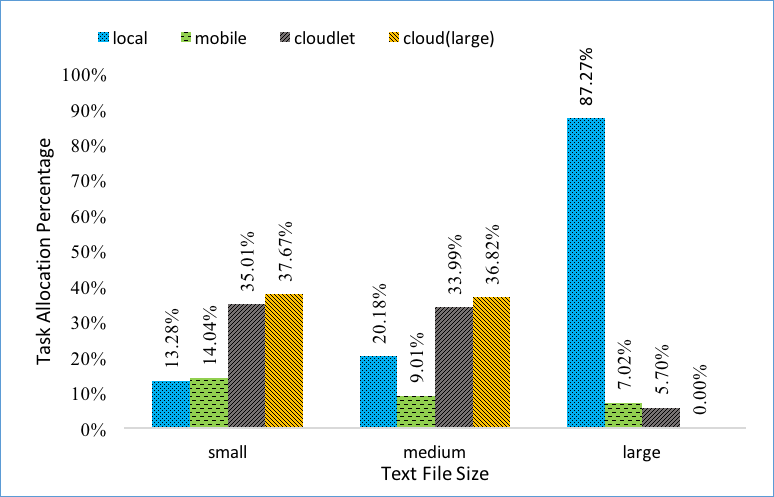}\par 
    \includegraphics[width=\linewidth]{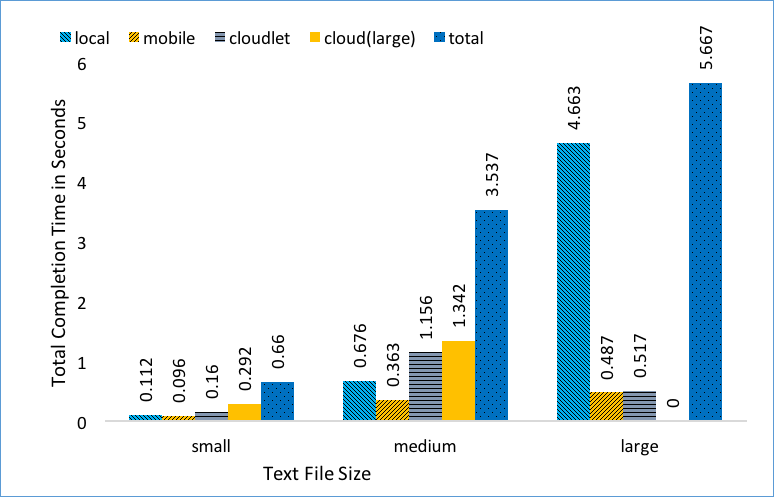}\par 
    \end{multicols}
\caption{Multisite Partial Offloading - The decision engine partitions the task for the local mobile device and multiple offloadees}
\end{figure*}

\begin{table*}[!t]
\centering
\label{alltests}
\caption{Experimentation Results}
\begin{multicols}{2}
 \includegraphics[width=\textwidth]{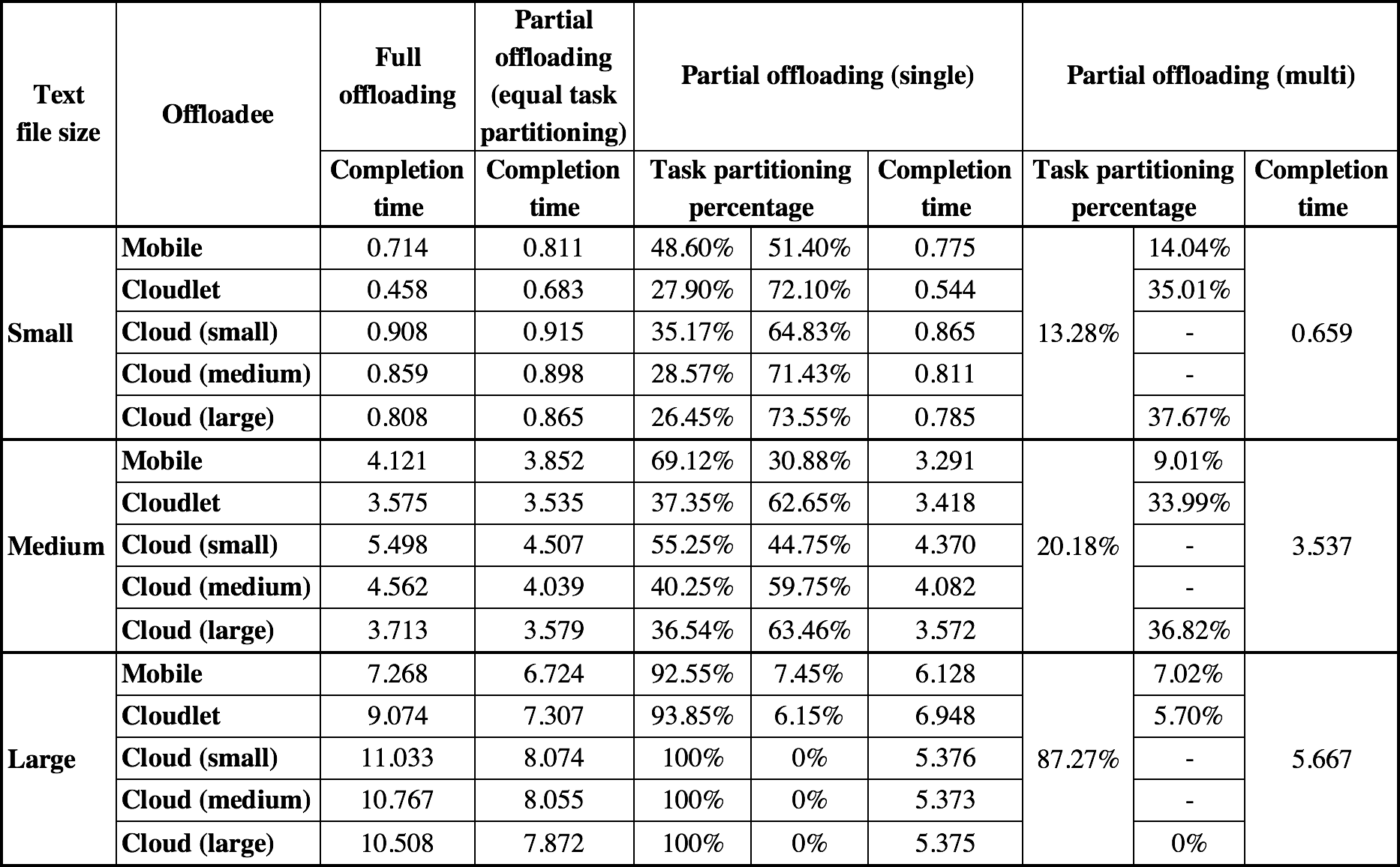}
\end{multicols}
\end{table*}

\section{Results and discussion}
It is already shown in the literature that offloading does not always benefit the lower end devices \cite{5445167}. In our first execution scenario, we send the text file to the destination and perform the search operation in the corresponding computation device. Despite few millisecond performance gains in the case of small text file, local execution is preferred to full offloading for medium and large text files. The partial offloading with equal partitioning of tasks among the local and external devices perform better in terms of reducing the overall network overhead occurrence in the previous execution scenario.

The single-site partial offloading produces better results than equal task distribution scenario in all of the offloading modes. It even produces a lower overall completion time for the cloudlet than the multi-site offloading scenario. This is due to the limited number of nearby mobile devices and cloudlets that are used in our testbed. Even though cloudlet-only full offloading performed better in both small and medium text file size execution scenarios, large text file size offloading was more efficient in MAMoC (9.074 vs. 5.667 seconds) as a larger portion of execution is performed locally. This is mainly due to the delay caused by data transfer in full offloading execution. For the large text file size offloading scenarios, the network overhead occupies a much larger portion of the total completion time than the time taken for the task execution. This shows a clear advantage of local execution over offloading when the needed computation is not as much. In short, the more computation and less data transfer is needed, offloading has more advantage.

Since we use Wi-Fi in the evaluation, the time of sending files and receiving results has an impact on the overall performance, but if cellular networks or Bluetooth are used, the offloading time will surely increase.

\section{Conclusions and Future Work}

In this paper, we presented an adaptive multisite offloading framework that takes into consideration dynamic context changes in MCC environment and offloads computation to multiple offloadees including nearby mobile devices, cloudlets, and remote cloud servers. We evaluated the proposed framework, and results showed that it can provide suitable offloading decisions based on the current context of the local device and the external platforms. We developed a text search application and conducted experiments with three different text file sizes on different offloading scenarios. Our results present different insights into the factors that affect the offloading decision.
Our future work includes increasing context parameters and the number of nearby devices and cloudlets in our experimental testbed. Our future experiments will not be confined to Wi-Fi and would explore other protocols such as Bluetooth Low Energy for nearby device-to-device communications and 4G for the mobile device and remote cloud communications. Moreover, we have built our framework with the assumption that the mobile user (offloader) stays in the same zone within the computation offloading duration. This cannot be guaranteed if the user moves away before the result from the destination is returned. The experiments in this paper were performed in a controlled setting. Enabling user mobility by live migrating the host mobile cloud containers is another future endeavor. An active research challenge is how to implement live migration in mobile cloud systems. 

%We have used WiFi throughout the experiments that we performed to evaluate the framework. In the future experiments, we would also like to use .

% Furthermore, we will be making the source code of the framework available online in future experiments, to further prove the soundness and transparency of our solution

\section{References}
\renewcommand\refname{}
\bibliographystyle{IEEEtran}
\bibliography{IEEEabrv,bare_conf}

\end{document}